\documentclass[11pt]{article}
\usepackage{amssymb,amsmath}
\usepackage{epsfig}
\usepackage{hyperref}
\usepackage{cite} 

\usepackage{times}

\newcommand{\qedsymb}{\hfill{\rule{2mm}{2mm}}}
\newenvironment{proof}[1][]{\begin{trivlist}
\item[\hspace{\labelsep}{\textbf\noindent Proof#1:\/}] }{\qedsymb\end{trivlist}}

\newcommand{\ignore}[1]{}

\newtheorem{theorem}{Theorem}[section]

\newtheorem{definition}[theorem]{Definition}
\newtheorem{observation}[theorem]{Observation}

\newtheorem{lemma}[theorem]{Lemma}

\newtheorem{coro}[theorem]{Corollary}

\newcommand{\smfrac}[2]{\mbox{$\frac{#1}{#2}$}}

\newcommand{\ket}[1]{|#1\rangle}
\newcommand{\bra}[1]{\langle#1|}

\newcommand{\ketbra}[2]{|#1\rangle\langle#2|}
\newcommand{\braket}[2]{\langle {#1} | {#2} \rangle}

\newcommand{\QMA}{{\sf{QMA}}}

\def\calH{{\cal{H}}}
\def\calS{{\cal{S}}}
\def\calT{{\cal{T}}}

\def\mns{{\mbox{-}}}

\newcommand{\mysymbol}[1]{{\mbox{\raisebox{-0.3em}{\includegraphics[height=5mm]{#1}}}}}

\newcommand{\xra}{\mysymbol{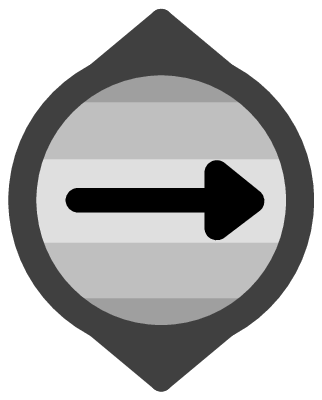}}
\newcommand{\xla}{\mysymbol{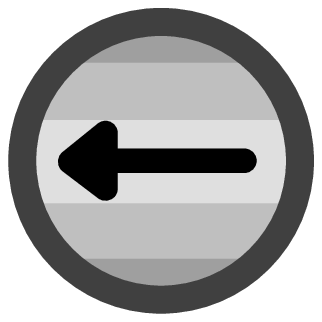}}
\newcommand{\xup}{\mysymbol{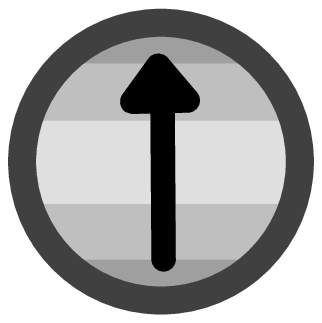}}
\newcommand{\xdo}{\mysymbol{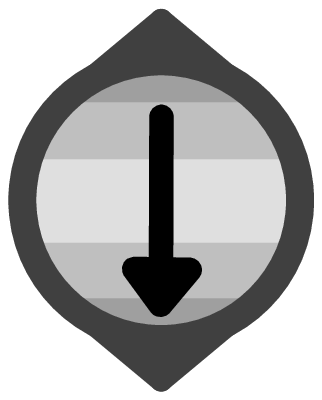}}
\newcommand{\xdra}{\mysymbol{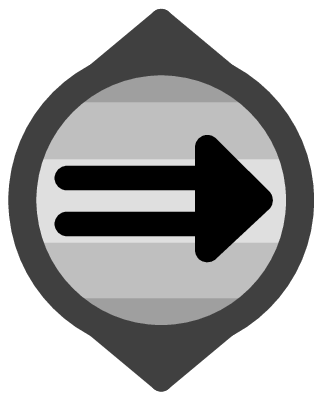}}
\newcommand{\xdup}{\mysymbol{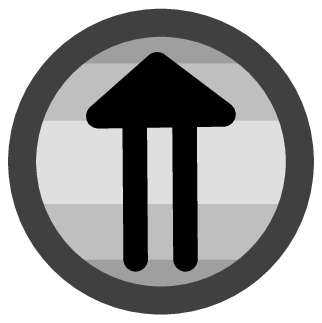}}

\newcommand{\Estate}{\mysymbol{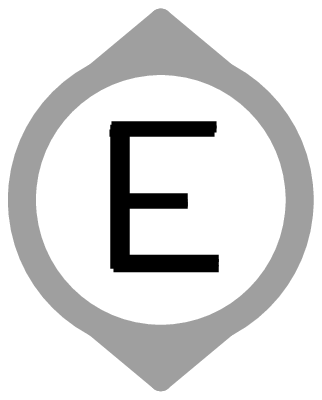}}
\newcommand{\estate}{\mysymbol{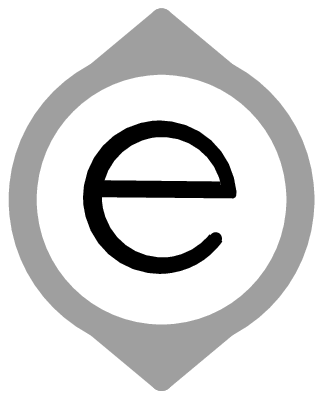}}
\newcommand{\Ustate}{\mysymbol{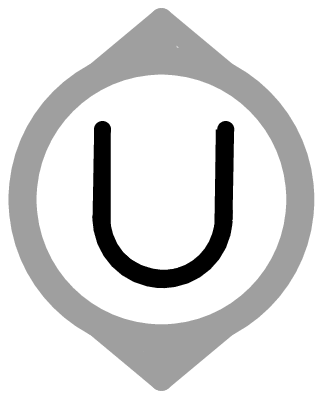}}
\newcommand{\ustate}{\mysymbol{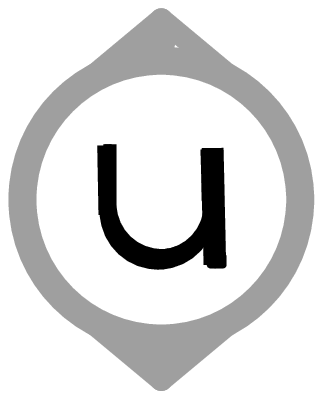}}
\newcommand{\Wstate}{\mysymbol{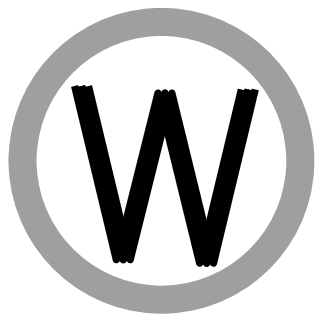}}
\newcommand{\wstate}{\mysymbol{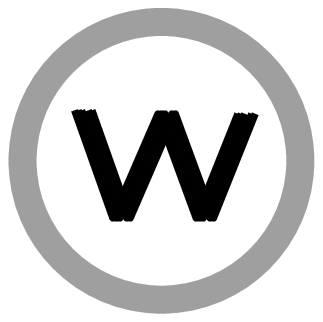}}

\newcommand{\blankState}{\mysymbol{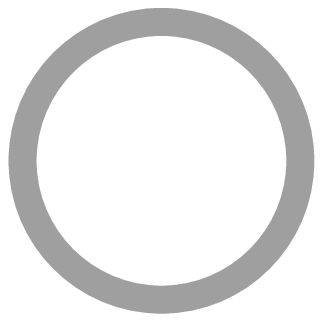}}

\newcommand{\leftend}{\mysymbol{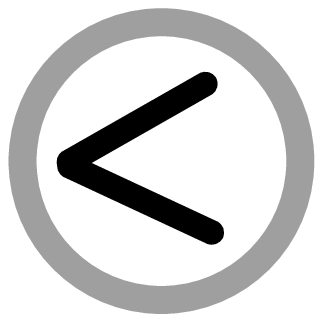}}
\newcommand{\rightend}{\mysymbol{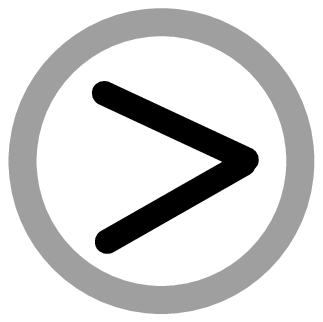}}

\newcommand{\focsv}[1]{{}}

\begin{document}

\title{Ground State Entanglement in One Dimensional 
Translationally Invariant Quantum Systems}

\date{\today}

\author{
Sandy Irani\thanks{E-mail: irani@ics.uci.edu.  Partially
supported by NSF Grant CCR-0514082  and CCF-0916181. Part of this work was done
while the author was visiting the Institute for Quantum Information at Caltech.}\\Computer Science
Department\\University of California, Irvine, USA
 }

\maketitle

\begin{abstract}
We examine whether it is possible for one-dimensional 
translationally-invariant Hamiltonians to have ground 
states with a high degree of entanglement.
We present a family of translationally invariant Hamiltonians $\{H_n\}$
for the infinite chain. The spectral gap of $H_n$ is $\Omega(1/poly(n))$.
Moreover, for any state in the 
ground space of $H_n$ and any $m$, there are regions of size $m$ with entanglement
entropy $\Omega(\min\{m,n\})$. A similar construction yields 
translationally-invariant Hamiltonians for finite chains 
that have unique ground states exhibiting high entanglement.
The area law proven by Hastings ~\cite{Hastings07}
gives a constant upper bound on
the entanglement entropy for 1D ground states
that is independent of the size of the region but
exponentially dependent on $1/\Delta$, where $\Delta$ is the
spectral gap.
This paper provides
a lower bound, showing a family of Hamiltonians for which the 
entanglement entropy scales polynomially with $1/\Delta$.
Previously, the best known such bound was logarithmic in $1/\Delta$.
\end{abstract}

\section{Introduction}

\vspace{-.1in}

Understanding and quantifying
entanglement in quantum systems is a central theme in
quantum information science. On one hand quantum entanglement is a
valuable resource that enables novel computation and communication.
On the other hand, the fact that some quantum systems have bounded
entanglement accounts for the success of computational methods such as DMRG
in finding ground states and simulating dynamics
\cite{White92,White93,DMRG-overview,vidal1, vidal2}.
We examine one dimensional quantum systems and ask what is the minimal
set of properties a system must have in order to exhibit a high degree
of ground state entanglement. In particular, do symmetries such as
translational invariance limit entanglement?

We present two closely related constructions. The first is
a single Hamiltonian term which acts on two 21-dimensional particles.
When the term is apllied to each neighboring pair of particles in  a
chain of $n$ particles, the resulting Hamiltonian
has a unique ground state and  a spectral gap of $1/poly(n)$.
We show that
 the entropy of the ground state
when  traced down to a linear number of particles on 
either end of the chain scales linearly with $n$.
If we allow ourselves boundary conditions in the form of a different
single-particle
term applied to the first and last particles of the chain,
the Hamiltonian is frustration-free.
 The boundary conditions can be removed with an additional term
applied to all the particles in the chain. With the additional term,
the resulting Hamiltonian is no longer frustration free.

The second construction is a family of 
translationally invariant Hamiltonians $\{H_n\}$.
When applied to any cycle whose size is
a multiple of $n$ or an infinite chain of particles,
the spectral gap is $1/poly(n)$.
The ground state is no longer unique, but for any state in the
ground space and any $m$, a constant fraction of the intervals of
length $m$ have entanglement entropy that is $\Omega(\min\{m,n\})$.
Moreover, there exists a state in the ground space which is translationally invariant
and has the property that
 every interval of length $m$ has entanglement entropy which is
$\Omega(\min\{m,n\})$. Nagaj has shown a way to reduce the number
of particles to $14$ \cite{Nag08}.

This paper builds on recent work examining the computational power
of one dimensional quantum systems in which it is shown that it is
possible to perform universal 
adiabatic quantum computation using a 1D quantum system \cite{aharonov-2009-287}.
In addition, it is shown that the problem of determining the ground
state of a one dimensional quantum system with nearest neighbor interactions
is $\QMA$-complete. 
Both results make critical
use of position-dependent terms in the Hamiltonian.
The intuition that symmetry in quantum systems is 
both  natural and might lead to more tractable systems,
has lead researchers to examine whether this construction
can be made translationally invariant.
For instance,
\cite{NWtranslation} gives a $20$-state translation-invariant modification of 
the
construction (improving on a $56$-state construction by \cite{JWZtranslation})
that can be used for universal 1-dimensional adiabatic computation.
These modifications require that the system be initialized to a particular
configuration in which the state of
each particle encodes some additional
information. 
The terms of the Hamiltonian, although identical, act differently
on different particles depending on their initial state. The ground state is
therefore degenerate and one determines which ground state
is reached by ensuring that the system starts in a particular state.
Liu et.~al.~\cite{LCV} show that the $N$-representability problem on fermions
is $\QMA$-complete. Since the states occupied by fermionic systems
are anti-symmetric, all two-particle
reduced density matrices are identical. However, the dimension of  the two-particle
matrices can grow polynomially with $N$, whereas we consider a constant number of
states per particle.
Kay~\cite{Kaytranslation} gives a construction showing that 
determining the ground energy of a one dimensional nearest-neighbor
Hamiltonian is $\QMA$-complete 
even with all two-particle terms identical, though the construction 
requires position-dependent one-particle terms. 
It is not clear how one would eliminate both the varying terms on
individual particles and the degeneracy of the ground state
and still obtain a complexity result. In particular, it is not
clear how a single Hamiltonian
term with 
bounded precision acting on a pair of particles with constant dimension 
would  encode a circuit or an input instance. 
However, it is still natural to  examine ground state entropy under
these limitations.

This work also relates to the area law for one dimensions proven
by Hastings \cite{Hastings07}. An area law asserts that in
ground states of local 
Hamiltonians, the entanglement entropy of the reduced state of a subregion
should scale with the boundary of the region as opposed to the volume of
the region. 
In one dimensions, the bounding area of a contiguous region is comprised only
of the two endpoints of the segment, so the area law says that the entropy of
entanglement should be independent of the size of the region.  The area law 
for one-dimensional systems proven by Hastings
depends exponentially on $1/\Delta$, where $\Delta$ is the spectral gap.
Gottesman and Hastings raised the question as to
whether this dependence on $\Delta$ is tight and towards this end
gave a family of Hamiltonians
on the infinite chain whose entanglement entropy 
scales as $\Omega((-\Delta \log \Delta)^{-1/4})$. Previously
studied systems have the property that the entropy of all intervals is
bounded by a constant
times $\log(1/\Delta)$. 

The results presented in this paper
independently provide a lower bound of this kind, although
the two sets of results have different features resulting from
the different motivation of the authors.
The \cite{GH08} construction is not translationally invariant 
as this is not required for the area law.
However, it is simpler, uses fewer states and 
the lower bound on entanglement as a function of the spectral gap is
a larger polynomial
($\Omega((-\Delta \log \Delta)^{-1/4})$ as opposed to
$\Omega(( \Delta)^{-1/12})$). Much of the effort in the construction
presented here stems from designing a translationally
invariant Hamiltonian.
In both constructions, the ground state achieves high entanglement on
some but not all of the regions. Technically, this is valid for a lower
bound on the area law since an area
law must give an upper bound on the entanglement entropy for all regions.
Nonetheless, Gottesman and Hastings point out
that their construction can be augmented,
using 81 instead of 9 states, 
to produce a ground state with high entanglement on every 
sufficiently large region.
Note that high entanglement entropy means polynomial in $1/\Delta$.
The entropy will not grow as the region size tends towards infinity
as this would violate Hastings' upper bound.
Finally there is the fact that the ground space for the construction
presented here is degenerate.
Note that this degeneracy is fundamentally different from the degeneracy
in the constructions of \cite{NWtranslation} 
and \cite{JWZtranslation} discussed above in that
every state in the ground space 
exhibits the desired entanglement properties.
There does not seem a way to break the degeneracy in this particular
construction using local,
translationally invariant rules. The basis of the ground space
consists of states which are translations of each other along the infinite chain.
Since the basis states for the ground space of  $H_n$ are periodic with 
period $n$, the dimension of the ground space
is $n$.
If one is willing to forego translational invariance, we
could use the construction for the finite chain presented
here to design a Hamiltonian for the infinite chain with
a unique ground state 
by simply repeating  copies of the Hamiltonian
for a chain of length $n$,
side by side. It should be noted that Hastings' upper bound is
only proven for Hamiltonians with degenerate ground states.
It is not clear whether the proof can be extended when the
degeneracy of the ground state is polynomial in $1/\Delta$,
where $\Delta$ is the difference between the lowest and second
lowest eigenvalues of the Hamiltonian.

In the constructions presented here and in \cite{GH08}, each Hamiltonian depends on the
parameter $n$
which in turn gives a
fixed upper bound for the entanglement entropy that
can be achieved for any region. It is unclear whether there is a way
to achieve entanglement entropy that is linear in the region size 
simultaneously for
all region sizes.
In the only known
examples of 1D ground states for which the entanglement entropy grows
asymptotically with the size of the region, the entropy depends logarithmically on
the region size \cite{vlrk03,ECP08}.
It has been conjectured 
that for any translationally invariant pure state, the
entropy of entanglement of a contiguous set of $n$ particles is sublinear as
$n$ grows \cite{fz05}.
This conjecture does not require that the state be a ground state of
a Hamiltonian (translationally invariant or otherwise). It is simply a question
about what can be achieved by a quantum state. In this sense it should be easier
to achieve high entanglement. On the other hand, the conjecture requires that the state itself
be translationally invariant.

\vspace{-.15in}

\subsection{Outline of Techniques}

\vspace{-.05in}

We begin with an overview of the construction for the finite chain,
much of which will also be used for the cycle and infinite chain.
We will have $21$ states on each site consisting of 2-state subsystems
which hold a qubit of data and 1-state subsystems. 
We use the term {\em configuration} to refer to an arrangement of the
states on a line without regard to the value of the data stored in the
qubit subsystems. 
The Hamiltonian applied to each pair of particles
will consist of a sum of terms of which there are two types.
The first type will have the form
$\ket{ab}\bra{ab}$ where $a$ and $b$ are single particle states.
We call these {\em illegal pairs} as it
has the effect of adding an energy penalty to any
state which has a particle
in state $a$ to the immediate left of a particle in state $b$.
We will say a configuration is {\em legal} if it does not contain
any  illegal pairs.
The second type of term will have the form:
$\frac{1}{2}
(\ketbra{ab}{ab} + \ketbra{cd}{cd} -
\ketbra{ab}{cd} - \ketbra{cd}{ab})$.
These terms enforce that for any eigenstate with zero
energy, if there is a configuration $A$ with two neighboring
particles in states $a$ and $b$, there must be a configuration $B$ with
equal amplitude that is the same as $A$ except that $a$ and $b$ are
replaced by $c$ and $d$.
Even though these terms are symmetric, we associate
a direction with them by denoting them as $ab \rightarrow cd$. 
These terms will be referred to as
{\em transition rules}. We will say that configuration
$A$  transitions into configuration $B$ by rule $ab \rightarrow cd$
if $B$ can be
obtained from $A$ by replacing an occurrence of $ab$ with an occurrence of
$cd$. We say that the transition rule
applies to $A$ in the forward direction
and applies to $B$ in the backwards direction. 
We will choose the terms so that for any legal configuration, at most one
transition rule applies to it in the forward direction and at most one
rule applies in the backwards direction. Thus, a ground state consists of
an equal superposition of
legal configurations such that there is exactly one transition rule
that carries each configuration to the next.
So far what we have described is a standard procedure in $\QMA$-completeness
results with the chain of configurations in the ground state corresponding
to the execution of the circuit through time \cite{Kitaev:book,KempeKitaevRegev06,
ad1,Oliveira:05a}.
For a one dimensional system, we have a small set of designated  states
called {\em control} states
and we enforce that any legal configuration has exactly one particle in a control 
state.
The transition rules apply only to the control state and a
particle to the immediate left or right, possibly moving the control state
left or right by one position, much like the head of a Turing Machine.
This idea was also employed in \cite{aharonov-2009-287}.

In the construction we present here, particles 
on the left half of the chain
start in a $2$-state subsystem that each encode a qubit in
state $\ket{+}$.
The control state will act as a courier, first getting entangled with
a qubit on the left end, and then moving to the right part of  the chain.
There it gets swapped with a qubit state, creating an entangled
qubit pair which spans  the center of the chain.
In each round trip made by the control
state, the number of entangled pairs increases by one and eventually
the number of entangled pairs spanning the center of the chain
is roughly half the number of particles in the chain.
When the qubit value of a particle on the left has been recorded by 
(or entangled with) the qubit value stored with the 
control particle, it transitions to a  
two-state subsystem which we represent by the symbol $\estate$. Similarly when the
particle on the right becomes entangled with the
qubit value of the control state, it 
transitions to a  two-state subsystem represented by the symbol
$\Estate$.  Thus the  particles in these states
build up over time on the two ends of the chain.
The transition rules ensure that the courier
changes direction as soon as it hits a  particle
in state ~$\estate$ or $\Estate$. The process is illustrated
below with overbrace spanning the newly created entangled
pair. In the actual construction more particle states will be
required to ensure that the process proceeds as depicted.

\begin{center}
\begin{tabular}{c}
\leftend \xdup\blankState \blankState \blankState \blankState \rightend \\
\leftend  $\overbrace{\estate \xra}$ \blankState \blankState \blankState \rightend  \\
\leftend  $\overbrace{\estate \blankState \xra}$ \blankState \blankState \rightend  \\
\leftend  $\overbrace{\estate \blankState \blankState \xra}$ \blankState \rightend \\
\leftend  $\overbrace{\estate \blankState \blankState \blankState \xra}$ \rightend \\
\leftend  $\overbrace{\estate \blankState \blankState \blankState \xdo}$ \rightend   \\
\leftend  $\overbrace{\estate \blankState \blankState \xla \Estate}$ \rightend    \\
\leftend  $\overbrace{\estate \blankState \xla \blankState \Estate}$ \rightend    \\
\leftend  $\overbrace{\estate \xla \blankState \blankState \Estate}$ \rightend   \\
\end{tabular}
\end{center}

Once the construction is given in detail, it is fairly evident
that it results in a high entanglement ground state and the main difficulty
is to establish that this ground state is unique. Thus, additional constraints
are required to give energy penalties to configurations that deviate from
this plan.
As was the case with 
the one dimensional $\QMA$-completeness construction of \cite{aharonov-2009-287}, 
we are not able to eliminate every undesirable configuration with
local checks and we need to show that some bad configurations are ruled out because
they
must eventually evolve (via forward or backwards transitions) to a configuration
which can be eliminated by local constraints.
For the problem addressed here, we need some means of enforcing that entangled pairs
actually span the center of the chain instead of spanning some boundary to the
far left or the far right as this could severely limit the number of entangled pairs.
We also need to enforce the condition that particles initially storing the $\ket{+}$
state to be entangled with other particles further down the chain
occur on the left half of the chain.
This could be easily managed with different terms on the left half
and the right half of the chain.
However, since we insist on uniform terms, we enforce these conditions by showing that
violating states will evolve   to  illegal states.
For example, if the number of particles in  state $\estate$ on the left
 is less than the number of particles in state $\Estate$ on the right,
we show that this
state will evolve via backwards transitions to a
 state with a collection of  particles in  state $\Estate$ on the right
and no particles in state $\estate$ on the left.
The first and last particles in the chain will be in special delimiter states
(with $\leftend$ at the left end and $\rightend$ at  the right end) which
will be used to detect this occurrence and trigger an energy penalty. 
This raises a new problem of how to make sure only the end particles are in these
delimiter
states. This is done by adding a penalty for any particle that is in a state
which is different from one of the delimiter states. Finally, we add even greater penalties
for any pair of the form $X \leftend$ or $\rightend X$
which ensures that only
the leftmost particle will be in state $\leftend$ and only the rightmost particle
will be in state $\rightend$.

The construction for the finite chain makes use of the fact that the
endpoints of the chain have only one neighbor. When we move to the cycle we not
longer have these special particles. 
We change the Hamiltonian for the cycle by
allowing the pair $\rightend \leftend$ with the effect that
the set of legal states 
become sequences of segments bracketed on either side. The legal
states look like
the following type of sequence wrapped around the cycle:
$$\leftend \cdots \rightend \leftend \cdots \rightend \leftend \cdots \rightend \leftend \cdots \rightend $$
Suppose we fix the locations of the $\leftend$ and $\rightend$ sites and consider the
space of states with those locations fixed. The Hamiltonian $H_n$ will be closed on that space
which allows us to analyze every such subspace separately.
Finally a term is added that  gives an energy penalty if there is a sequence
from a $\leftend$ site to a $\rightend$ site whose length is
not exactly equal to $n$. Thus, the ground state for a cycle of length $tn$ will be
$t$ copies of the ground state for a finite chain of length $n$
tensored together. There are $n$ such global states, each a rotation of the others.
We will show that the bounds on the spectral gap and the entanglement entropy
are independent of $t$, so as $t$ goes towards infinity, the bounds will
still hold which means that they also hold for the infinite chain.

\section{The Basic Construction on a Finite Chain}

The $21$ states
in each site consist of $2$-state subsystems (different versions of
a qubit holding data), represented by elongated shapes (e.g.,
\xra), and $1$-state subspaces, represented by round shapes (e.g.,
\xla). 
\footnote{This notation, which has been adapted for
the construction presented here, was developed in \cite{aharonov-2009-287} 
in collaboration with Oded Regev.}
Three of the $2$-state subsystems and three of the $1$-state subsystems
will be control sites, which will be
represented by dark shapes and can be thought of as pointers on the
line that trigger transitions. Light-colored shapes represent
a site that is inactive, waiting for the active site to come nearby.
There will only be one control site in any legal configuration.  
Particles in states denoted by lower case letters will 
always be to the left of the control site and particles in states denoted
by upper case letters will be to the right of the control state.
When needed, we will indicate the value of
the qubit stored in a $2$-state subsystem with a subscript indicating the state,
such as $\xra_+$ or $\xra_1$. 
We
have the following types of states:
\medskip

\begin{tabular}{ll}
 \underline{Inactive states}  & \underline{Control states} \\ 
  \Estate, \estate: Qubits entangled with another site & \xra, \xdra:  Right-moving control states \\
\Ustate, \ustate: Qubits unentangled with another site  & \xla :  Left-moving control state  \\
  \Wstate, \wstate: Particles waiting to be entangled &  \xup, \xdup : Left-end turning control states\\
  \leftend, \rightend : The left and right end delimiters & \xdo: Right-end turning control state\\
\end{tabular}

\vspace{.1in}

We start by introducing the set of transition rules.
Unless otherwise specified, a rule applied to a $2$-state subsystem
is summed over all possible values for the  qubits, with  the control
particle keeping its value and the non-control particle  keeping its value.
For example, the rule $\xra \Ustate \rightarrow \ustate \xra$
would be  the sum of $\xra_x \Ustate_y \rightarrow \ustate_y \xra_x$
over all possible values for $x,y \in \{0,1\}$.
The exceptions to this are made explicit in the set of rules below.
The sum of all the resulting terms is denoted by $H_{trans}$.

The rules involving sites with single arrows
are used throughout most of the evolution of
the configurations. Rules involving sites with double arrows 
occur only during the first iteration of the construction
and are used to check the validity of the starting configuration.

\vspace{.1in}
\noindent
\underline{{\textbf Transition Rules:}}
\begin{enumerate}
\item $\xra \Ustate \rightarrow \ustate \xra$,
$\xra \Wstate \rightarrow \wstate \xra$: Sweeping to the right past $\Ustate$ and $\Wstate$
sites,
transforming them to $\ustate$ and $\wstate$.
\item $\xra \Estate \rightarrow \xdo \Estate$: Control turns once an entangled
site  is  reached on the right end.
\item $\wstate \xdo_x \rightarrow \xla \Estate_x$: Control starts moving left and
transfers its qubit state to the $\Estate$.
\item $\wstate \xla \rightarrow \xla \Wstate$, $\ustate \xla \rightarrow \xla \Ustate$:
Control sweeps to the left past $\ustate$ and $\wstate$ sites, transforming them to
$\Ustate$ and $\Wstate$ sites. 
\item $\estate \xla \rightarrow \estate \xup$: Control turns once an entangled
site is reached on the left end.
\item $\xup \Ustate_x \rightarrow \estate_x \xra_x$: Control starts moving right. Qubit values of
$\estate$ and $\xra$ become entangled.
\item $\xdup \Ustate_x \rightarrow \estate_x \xdra_x$: Control starts moving right. Qubit values of
$\estate$ and $\xdra$ become entangled.
\item $\xdra \Ustate \rightarrow \ustate \xdra$,
$\xdra \Wstate \rightarrow \wstate \xdra$: Sweeping to the right past $\Ustate$ and $\Wstate$,
transforming them to $\ustate$ and $\wstate$.
\item $\xdra \rightend \rightarrow \xdo \rightend$: Control turns once the right
end delimeter is reached.
\end{enumerate}

Rules 6 and 7 above are the rules that create the entanglement in
the construction. Later in the construction, we will add terms which
enforce that for any state in the support of a ground
state, if a particle is in state \Ustate, then the value
stored in the qubit of that 2-state subsystem must be $\ket{+}$.
The action of transition rules 6 and 7 on a state of this kind 
create an entangled pair:

$$\xup \left( \frac{1}{\sqrt{2}} \Ustate_0 + \frac{1}{\sqrt{2}} \Ustate_1 \right) 
\rightarrow \frac{1}{\sqrt{2}} \left( \estate_0 \xra_0 + \estate_1 \xra_1 \right).$$
$$\xdup \left( \frac{1}{\sqrt{2}} \Ustate_0 + \frac{1}{\sqrt{2}} \Ustate_1 \right) 
\rightarrow \frac{1}{\sqrt{2}} \left( \estate_0 \xdra_0 + \estate_1 \xdra_1 \right).$$

We wait to introduce the rule that enforces that a \Ustate ~particle must be in the
state $\Ustate_+$ because it will be convenient for the first part of the
proof to work with standard basis states.
A state in the standard basis is first specified by its configuration and then
by a $0$ or $1$ value for each $2$-state subsystem in the  configuration.
Thus, a standard basis state is represented by a string of symbols from the set
\begin{equation}
\label{eq:standardstates}
\Estate_0, \Estate_1, \estate_0, \estate_1, \Ustate_0, \Ustate_1, \ustate_0, \ustate_1, \Wstate, \wstate,
\xra_0, \xra_1,
\xdra_0,\xdra_1, \xla, \xup, \xdup, \xdo_0, \xdo_1, \leftend, \rightend.
\end{equation}
A configuration is represented by a string of symbols from this set without the $0$ and $1$
subscripts specified.
The transition rules and illegal pairs are all specified with respect to
standard basis states.

We are now ready to describe the evolution of configurations in the
target ground state. We assume that we start with a configuration
in the following form
$\leftend \xdup \Ustate^m \Wstate^m \rightend$,
where $n= 2m+3$.
(We will eventually prove that the low energy states exist only when $n$ is odd).
The construction is illustrated with
a small example below to show what happens as each rule is applied:

\begin{center}
\begin{tabular}{cc}
Round One & Round Two\\
\leftend $\xdup$ \Ustate \Ustate \Wstate \Wstate \rightend &  \leftend  \estate \xup \Ustate \Wstate \Estate \rightend \\
\leftend  \estate \xdra \Ustate \Wstate \Wstate \rightend  & \leftend  \estate \estate \xra \Wstate \Estate \rightend \\
\leftend  \estate \ustate \xdra \Wstate \Wstate \rightend  &  \leftend  \estate \estate \wstate \xra \Estate \rightend \\
\leftend  \estate \ustate \wstate \xdra \Wstate \rightend &  \leftend  \estate \estate \wstate \xdo \Estate \rightend \\
\leftend  \estate \ustate \wstate \wstate \xdra \rightend &  \leftend  \estate \estate \xla \Estate \Estate \rightend \\
\leftend  \estate \ustate \wstate \wstate \xdo \rightend   &  \leftend  \estate \estate \xup \Estate \Estate \rightend \\
\leftend  \estate \ustate \wstate \xla \Estate \rightend   & \\
\leftend  \estate \ustate \xla \Wstate \Estate \rightend   & \\
\leftend  \estate \xla \Ustate \Wstate \Estate \rightend  &  \\
\end{tabular}
\end{center}

Now we describe a set of terms that 
are designed to
ensure that the state corresponding to the evolution of configurations
shown above is the only low energy state. The constraints are expressed in terms of
illegal pairs (pairs of states which cause an energy penalty if they
appear side by side in a configuration). 
It is sometimes convenient to describe a set of states informally
such as (UPPERCASE) which denotes any state represented  by an upper case letter.
The label for each set is indicated below along with the set of states
in that set.

\begin{center}
\begin{tabular}{|c|c|}
\hline
(anything) & the set of all states \\
\hline
(lowercase) & $\{ \estate, \ustate, \wstate \}$\\
\hline
(UPPERCASE,\rightend) & $\{ \Estate, \Ustate, \Wstate,\rightend \}$\\
\hline
(lowercase,\leftend) & $\{ \estate, \ustate, \wstate,\leftend \}$\\
\hline
(UPPERCASE) & $\{ \Estate, \Ustate, \Wstate \}$\\
\hline
{\textbf (control)} & $\{ \xra, \xdra, \xla, \xup, \xdup, \xdo \}$ \\
\hline
\end{tabular}
\end{center}

\vspace{.1in}
\noindent
\underline{{\textbf Illegal Pairs:}}
\begin{enumerate}

\item
\rightend (anything), (anything) \leftend: if $\rightend$ is in the system at all, must be at the right.
If $\leftend$ is in the system at all, it must be at the left end.
\item
(UPPERCASE)(lowercase), {\textbf (Control)} (lowercase),
(UPPERCASE) {\textbf (Control)} : Lower case sites before control sites before upper case sites.
\item
{\textbf (Control)} {\textbf (Control)} : At most one control site.
\item
(lowercase,\leftend) (UPPERCASE,\rightend):
 Lower case and $\leftend$ sites must be buffered from
upper case and $\rightend$ sites by a control site.

\item
\wstate \estate, \ustate \estate :  
$\estate$ sites before $\ustate$ and $\wstate$ sites.
\item
\Estate \Wstate, \Estate \Ustate : 
 $\Wstate$ and $\Ustate$ sites before $\Estate$ sites.

\item
\ustate\xup , \ustate\xdup , \wstate\xup ,  \wstate\xdup
: Control turns at the left end of unentangled and waiting sites.
\item
\xdo\Ustate  , \xdo\Wstate 
: Control turns at the right end of unentangled and waiting sites.

\item
\wstate \ustate , \Wstate \Ustate : $\ustate$ and $\Ustate$ sites should come before $\wstate$ and $\Wstate$ sites.
\item
\xdra \Estate , \xra \rightend  : Will be used to enforce the correct initial configuration.
\item
\leftend \xup, \estate \xdup, \xdup \Estate   : Will be used to enforce correct initial configuration.
\item
\xup \Wstate, \ustate \xdo, \estate \xdo : Will ensure that the number of sites in $\ustate$ and $\Ustate$ is same
as the number of sites in $\wstate$ or $\Wstate$.
\item
$\estate_0 \xra_1$, $\estate_1 \xra_0$ : Will ensure that pairs of qubits are properly entangled.
\item
$\estate_0 \xdra_1$, $\estate_1 \xdra_0$ : Will ensure that pairs of qubits are properly entangled.
\end{enumerate}

To define the set of terms arising from the illegal pairs as they act on states (and not
just $2$-state subsystems), 
we simply sum over all combinations of qubit values,
except for constraints in items $13$ and $14$ which are explicitly specified.
For example, the illegal pair
$\ustate \estate$ gives rise to 
the term $\ketbra{\ustate \estate}{\ustate \estate}$ which is then is expanded as follows:
$$\ketbra{\ustate_0 \estate_0}{\ustate_0 \estate_0} +
\ketbra{\ustate_0 \estate_1}{\ustate_0 \estate_1} +
\ketbra{\ustate_1 \estate_0}{\ustate_1 \estate_0} +
\ketbra{\ustate_1 \estate_1}{\ustate_1 \estate_1}.$$
The resulting term obtained from adding all the constraints above is $H_{legal}$.

Although we will ultimately insist that legal configurations do not contain any illegal pairs,
it will be convenient to work with a larger set of configurations/states which only 
omit illegal
pairs listed in items 1 through 8. 

\begin{definition}
A configuration is said to be {\textbf legal} if it has no illegal pairs listed in items $1$ through $12$.
(The illegal pairs in items $13$ and $14$ apply only to states).
A state is {\textbf legal} if it has no illegal pairs.
A configuration or state is {\textbf well-formed} if it does not contain any of the illegal pairs
listed in items 1 through 8.
\end{definition}

We start by characterizing the set of well-formed configurations.
In doing so, we will use the following notation:
$ \xra ^*$ will denote a sequence of sites in state $\xra$
of arbitrary (possibly zero) length. $( \xra / \xla / \xdra )$ is a single state which
is either $\xra$ or $\xla$ or $\xdra$. 
$\epsilon$ will denote an empty string of symbols.

\begin{lemma}
\label{lem:well-formed}
The set of well-formed configurations are those configurations which conform to one of the
expressions below or any substring of one of the expressions below:
$$(\leftend / \epsilon) \estate^* (\ustate / \wstate)^* ( \xra / \xla / \xdra )(\Wstate / \Ustate )^* \Estate^* (\rightend / \epsilon)$$
$$(\leftend / \epsilon) \estate^*  (\xup / \xdup) (\Wstate / \Ustate )^* \Estate^* (\rightend / \epsilon)$$
$$(\leftend / \epsilon) \estate^*  (\wstate / \ustate )^* \xdo \Estate^* (\rightend / \epsilon)$$
\end{lemma}

\begin{proof}
Constraint $1$ ensures that if  there is a $\rightend$, then it must be
the right-most particle in the chain. Similarly, if there
is a $\leftend$ then it is the left-most particle in the chain.
The remaining states are all either lower case, upper case or control states,
so well-formed states must be of the form
$$(\leftend / \epsilon)(\mbox{lowercase} / \mbox{\textbf Control} /\mbox{UPPERCASE})^*(\rightend / \epsilon).$$
Constrain $2$ says that lower case sites must precede control sites which must
preced upper case sites, so we have:
$$(\leftend / \epsilon)(\mbox{lowercase} )^*( \mbox{\textbf Control} )^*
(\mbox{UPPERCASE})^*(\rightend / \epsilon).$$
Constraint $3$ enforces that there can be at most one consecutive control state which yields:
$$(\leftend / \epsilon)(\mbox{lowercase} )^*( \mbox{\textbf Control} / \epsilon )
(\mbox{UPPERCASE})^*(\rightend / \epsilon).$$
Constraint $4$ says that if there are particles in a lower case or $\leftend$ state
and there are particles in a upper case or $\rightend$ state, then there must be
something to buffer them. This something can only be a control site because the
configurations are restricted as indicated above. Thus we know the configuration
must be a substring of:
$$(\leftend / \epsilon)(\mbox{lowercase} )^*( \mbox{\textbf Control})
(\mbox{UPPERCASE})^*(\rightend / \epsilon).$$
Constraints in item $5$ ensure that within the lower case sites,
$\estate$ must precede $\ustate$ and $\wstate$ sites.
Constraints in item $6$ ensure that within the upper case sites,
$\Ustate$ and $\Wstate$ sites must precede $\Estate$ sites. So a well-formed
configuration must be a substring of:
$$(\leftend / \epsilon) \estate^* (\ustate / \wstate)^* \mbox{(\textbf Control)}
(\Wstate / \Ustate )^* \Estate^* (\rightend / \epsilon)$$
If the control symbol is one of $\xra , \xla , \xdra$, there are
no further constraints. If the control symbol is $\xup$ or $\xdup$,
then constraint $7$ says that we have no $\ustate$ or $\wstate$ particles.
If the control symbol is $\xdo$, then constraint $8$ says that we have
no $\Ustate$ or $\Wstate$ particles.
\end{proof}

Any state that corresponds to a configuration that is not
well-formed will have an energy penalty from one of the terms from the first eight
items in the list of illegal pairs.
Thus, we can focus our attention on the well-formed states.
The following observation follows from an inspection of the rules.

\begin{observation}
\label{ob:closed}
The set of well-formed states is closed under the 
transitions rules in both the forward and the backward
directions.
\end{observation}

Observation \ref{ob:closed} and the following lemma
show that the transition rules 
are well behaved on the set of well-formed states.

\begin{lemma}
\label{lem:chain}
For each well-formed state, at most one transition rule
will apply in the forward direction and at most one will
apply in the reverse direction.
\end{lemma}

\begin{proof}
We use the fact that a well-formed state has at most
one site in a control state. Every transition rule,
whether applied in the forward or reverse direction, involves a control
site and a site to the immediate left or right.
Furthermore, the type of control state uniquely determines
whether it will be the site to the left or the right that it will
be involved in the transition in the forward direction.
The same is true for the reverse direction. 
\end{proof}

We now define a graph where each state in the standard basis is identified
with a node in the graph and there is a directed edge from one state to
another if there is a transition rule that takes one state to
the other. 
We will call this graph the {\textbf state graph} for our construction.
Observation \ref{ob:closed} implies that the
set of well-formed states is disconnected from the rest of the
states. Furthermore, by Lemma \ref{lem:chain}, the graph when restricted to the set of
well-formed states forms a set of disjoint directed paths. 
If there is a maximal path in the graph that has no illegal states, then
a uniform superposition over those states is a zero energy eigenstate.
Our next task is to characterize these paths.
We would like to be able to say that the zero eigenstates
are exactly those that correspond to the sequence of configurations 
illustrated earlier as our target ground state.
Unfortunately, this is not necessarily true.
For example, we could have a legal state which does 
not have a particle in a control
state at all and this state will correspond to a single isolated node component
of the graph. We can not enforce by local checks that a
state has a control state.
However, we will be able to make this assertion if we assume that the state begins
and ends with~ $\leftend$ and $\rightend$. Later we will need to add
terms to our Hamiltonian to ensure the endpoints of the chain are in
these delimiter states.


\begin{definition} A standard basis state is {\textbf bracketed} if the leftmost
particle is in state \leftend and the rightmost particle is in state
\rightend.
\end{definition}

Note that the transition rules do not alter the number or locations of
the ~$\rightend$ and $\leftend$ sites, so  the set of states in a path
in the state graph are either all bracketed or all not bracketed.
Thus, we can refer to a path as bracketed or not.
Now we have several definitions that we will use to
characterize the states in the target ground
state. The first definition enumerates a set of properties that guarantee that the
entangled pairs 
span the center of the chain.

\begin{definition}
A bracketed state  is said to be {\textbf balanced} if it is well-formed and
the following conditions hold:
\begin{enumerate}
\item
Every site in state ~$\Wstate$ or ~$\wstate$ occurs to the right
of every site in state ~$\Ustate$ or ~$\ustate$ in the chain.
\item If the control symbol is ~$\xdra$, $\xra$ or $\xdo$ then the number or
particles in state $\estate$ is one more than the number of particles in ~$\Estate$
and the number of particles in state ~$\Ustate$ or ~$\ustate$ is one less than the
number of particles in state ~$\Wstate$ or $\wstate$. 
\item
If the control symbol is ~$\xdup$, $\xup$ or $\xla$, then the number or
particles in state $\estate$ is equal to the number of particles in state
$\Estate$ and the number of particles in state ~$\Ustate$ or $\ustate$ is equal to the
number of particles in state $\Wstate$ or $\wstate$.
\item
If the control symbol is ~$\xdra$ there is one particle in state $\estate$ and
if the control symbol is ~$\xra$, there are at least  two particles in state $\estate$.
\item
If the control symbol is  ~$\xdup$ there are no particles in state $\estate$ and if
the control symbol is ~$\xup$ or ~$\xla$, there is at least one particle in state ~$\estate$.
\end{enumerate}
\end{definition}

The next definition will be used to ensure that the  ground
state is properly entangled. However, the property defined
 is not itself a quantum property in
that it is defined for standard basis states which are represented by strings
of symbols from the set of particle states specified in (\ref{eq:standardstates}).
The definition refers to the value of a qubit stored in a 2-state 
subsystem, but since we are referring to standard basis states,
each qubit value is always $0$ or $1$. 

\begin{definition}
Consider a balanced state in the standard basis with $r$ particles in state $\Estate$.
The state is {\textbf consistent} if for $i =1$ to $r$, the $i^{th}$
site in state ~$\Estate$ from the right  end has the same qubit value
as the $i^{th}$ state in ~$\estate$ from the left.
Furthermore, if the control symbol is ~$\xra$, $\xdra$ or $\xdo$, then
the qubit stored in the control state is the same as the qubit
stored in the rightmost site in state $\estate$.
\end{definition}

We will show that if a path in the state graph is composed of
bracketed, balanced and consistent nodes then the
first state in the path has the following configuration:
$\leftend \xdup \Ustate^m \Wstate^m \rightend$.
We say that any state that corresponds to this configuration is a {\em good start state}.
The next lemma says that if a state is bracketed, balanced and consistent, then
it belongs to a path whose initial state is a good start state. Furthermore the path is composed entirely
of legal states. Then in the following two lemmas,
we show that if a state is bracketed but not balanced or consistent,
it belongs to a path that has at least one illegal state. This will leave three
possibilities for a path: it is not bracketed, it 
it contains an illegal state, or it starts in a good start state and is composed
entirely of legal states.

\begin{lemma}
\label{lem:goodstates}
Consider a bracketed, balanced and consistent state and the path $p$ in the state
graph to which it belongs. The path $p$ contains only legal states. Furthermore, the start
state of $p$ is a good start state.
\end{lemma}

\begin{proof}
We will enumerate the possibilities for a bracketed, balanced and consistent state
and show that after a transition rule is applied in either the forward or reverse
direction, it remains bracketed, balanced and consistent. Furthermore, none of these
states contains an illegal pair.
We will also show that ~$\leftend \xdup \Ustate^m \Wstate^m \rightend$
is the only bracketed, balanced, consistent
configuration for which there is no
tranisition rule that applies in the reverse direction. This makes it the only
candidate for the first configuration in the path $p$.
Let $m = (n-3)/2$. We will refer to the sequence of
$\ustate$, $\Ustate$, $\wstate$ and $\Wstate$ particles as the 
{\em middle section}
We will break the argument down into cases, depending 
on the type of control symbol in the state:

\begin{description}

\item{ $\xdup$: } There is only one balanced configuration for this control
state which is
~$\leftend \xdup \Ustate^m \Wstate^m \rightend$. The only rule that applies to it
does so in the
forward direction and results in ~$\leftend \estate \xdra \Ustate^{m-1} \Wstate^m \rightend$. 
This is a balanced configuration. Since the rule entangles the $\estate$ qubit with the
$\xdra$, it is also consistent and legal.

\item{ $\xdra$: } The possible configurations are 
~$\leftend \estate \ustate^{j} \xdra \Ustate^{m-j-1} \Wstate^{m}  \rightend$,
where $0 \le j \le m-1$
 or
~$\leftend \estate \ustate^{m-1} \wstate^{j} \xdra \Wstate^{m-j}  \rightend$,
where $0 \le j \le m$.
If the control state is at the left end of the middle section and the state is consistent, 
it will transition in the reverse direction to
~$\leftend \xdup \Ustate^m \Wstate^m \rightend$. If the control state is 
at the right end of the middle section, it will
transition in the forward direction to
~$\leftend \estate \ustate^{m}  \wstate^{m} \xdo   \rightend$ .
Otherwise, when a  transition rule is applied  in the forward direction, the 
control state moves one site to the right and when a transition rule
is applied in the reverse direction, it moves one site to the left.
The state remains bracketed, balanced, legal and consistent.

\item{ $\xup$: }
The configuration must have the following form:
~$\leftend \estate^{i} \xup \Ustate^{m-i} \Wstate^{m-i} \Estate^{i} \rightend$,
where $1 \le i \le m$.
If $m=i$, there is no transition in the forward direction.
If $m<i$, in the forward direction it goes to
~$\leftend \estate^{i+1} \xra \Ustate^{m-i-1} \Wstate^{m-i} \Estate^{i} \rightend$.
The rule entangles the qubit values for the $\xra$ and the rightmost $\estate$,
so the state remains consistent. In the reverse direction, it goes to
~$\leftend \estate^{i} \xla \Ustate^{m-i} \Wstate^{m-i} \Estate^{i} \rightend$.
The resulting states are bracketed, balanced, consistent and legal.

\item{ $\xra$: }The first possible configurations is 
~$\leftend \estate^{i+1} \ustate^{j} \xra \Ustate^{m-j-i-1} \Wstate^{m-i} \Estate^i \rightend$,
where $1 \le i \le m-1$ and $0 \le j \le m-i-1$.
The second is
~$\leftend \estate^{i+1} \ustate^{m-i-1} \wstate^{j} \xra \Wstate^{m-j-i} \Estate^i \rightend$,
where $1 \le i \le m-1$ and $0 \le j \le m-i$.
If the control state is at theleft end of the middle section and the state is consistent, 
it will transition in the reverse direction to
~$\leftend \estate^i \xup \Ustate^{m-i} \Wstate^{m-i} \Estate^i \rightend$. 
If the control state is 
at the right end of the middle section, it will
transition in the forward direction to
~$\leftend \estate^{i+1} \ustate^{m-i-1}  \wstate^{m-i} \xdo  \Estate^{i} \rightend$.
Otherwise, when a  transition rule is applied  in the forward direction, the 
control state moves one site to the right and when a transition rule
is applied in the reverse direction, it moves one site to the left.
The resulting states are bracketed, balanced, consistent and legal.

\item{ $\xdo$: }
The configuration looks like
$\leftend \estate^{i+1} \ustate^{m-i-1}  \wstate^{m-i} \xdo  \Estate^{i} \rightend$,
for $0 \le i \le m-1$.
In the reverse direction, it transitions to
~$\leftend \estate^{i+1} \ustate^{m-i-1}  \wstate^{m-i} \xra  \Estate^{i} \rightend$.
In the forward direction, it transitions to
~$\leftend \estate^{i+1} \ustate^{m-i-1}  \wstate^{m-i-1} \xla  \Estate^{i+1} \rightend$.
The forward transition rule transfers the qubit value from the $\xdo$ state to the leftmost
$\Estate$, so it remains consistent. 
The resulting states are bracketed, balanced, consistent and legal.

\item{ $\xla$: }
The first possible configuration is
$\leftend \estate^{i} \ustate^{m-i}  \wstate^{j} \xla \Wstate^{m-j-i} \Estate^i \rightend$,
where $1 \le i \le m$ and $0 \le j \le m-i$.
The second
$\leftend \estate^{i} \ustate^{j} \xla \Ustate^{m-j-i} \Wstate^{m-i} \Estate^i \rightend$,
where $1 \le i \le m$ and $0 \le j \le m-i$.
If the control state is at the left end of the middle section and the state is consistent, 
it will transition in the forward direction to
$\leftend \estate^i \xup \Ustate^{m-i} \Wstate^{m-i} \Estate^i \rightend$. 
If the control state is 
at the right end of the middle section, it will
transition in the reverse direction to
$\leftend \estate^{i} \ustate^{m-i}  \wstate^{m-i+1} \xdo  \Estate^{i-1} \rightend$.
The $\xdo$ state 
takes the qubit value of the leftmost $\Estate$ that it replaces
and so 
the state remains consistent.
Otherwise, when a  transition rule is applied  in the forward direction, the 
control state moves one site to the left and when a transition rule
is applied in the reverse direction, it moves one site to the right.
The state remains bracketed, balanced, legal and consistent.

\end{description}
\end{proof}

\begin{lemma}
\label{lem:balanced}
If a state in the standard basis is bracketed and well-formed but not balanced, it will evolve eventually 
(via forward or backwards rules) to a configuration which is not legal.
\end{lemma}

\begin{proof}
Starting with the first condition on balanced configurations,
the only way for a configuration to have a $\Wstate$
or $\wstate$ to the left of a $\Ustate$ or $\ustate$ and not have an illegal
pair from item $9$, is 
to have \wstate(Control)\Ustate. 
Item $7$ forbids \xup or \xdup to the left of \wstate and item $8$ forbids
\xdo to the left of a \Ustate, so the control state
must be one of~ $\xra$, $\xdra$ or $\xla$. In the next step, the configuration
will transition to \wstate\ustate(Control) or (Control)\Wstate\Ustate which will
create illegal pair \wstate\ustate or \Wstate\Ustate from item $9$.

Now let's assume that the condition on the $\Estate$ and $\estate$ sites is violated.
We will first address the problem that there are too many $\estate$ sites.
This will eventually evolve backwards to a configuration that looks like
$\leftend \estate \ldots \estate  \ustate \ldots \ustate \wstate \ldots  \wstate \xdo \rightend$.
Transitioning in the backwards direction, the $\rightend$ site triggers the
control state to transition to $\xdra$ instead of $\xra$, resulting in
$\leftend \estate \ldots \estate  \ustate \ldots \ustate \wstate \ldots  \wstate \xdra \rightend$.
The $\xdra$ state will sweep leftwards in the reverse direction and
eventually hit the $\estate$ site resulting in
$\leftend \estate \ldots \estate \xdra  \ustate \ldots \ustate \wstate \ldots \wstate \rightend$
which will transition to
$\leftend \estate  \ldots \estate \xdup  \ustate \ldots \ustate \wstate \ldots  \wstate \rightend$,
creating illegal pair \estate \xdup from item $11$.

Similarly, if there are too many $\Estate$ sites, we will eventually transition backwards
to a configuration that looks like
$\leftend \estate \ustate \ldots \ustate \wstate \ldots \wstate \xdo \Estate  \ldots \Estate \rightend$.
This will transition to 
$\leftend \estate \ustate \ldots \ustate \wstate \ldots  \wstate \xra \Estate \ldots \Estate  \rightend$.
The $\xra$ state will sweep leftwards and eventually hit the $\estate$ resulting in 
$\leftend \estate \xra \Ustate \ldots \Ustate \Wstate \ldots \Wstate \Estate \ldots \Estate    \rightend$.
This transitions backwards to 
$\leftend \xup \Ustate \ldots \Ustate \Wstate \ldots  \Wstate \Estate \ldots \Estate   \rightend$
which again creates illegal pair \leftend\xup from item $11$.
We need to handle the configuration $\leftend \xdup \Estate \ldots \Estate \rightend$
separately because the $\xdup$ state does not have a transition in the reverse
direction. However, this configuration is disallowed because the~ $\xdup \Estate$ 
pair is one of the illegal states pairs in item $11$.

Now we will assume that the number of $\Estate$ sites and $\estate$ sites are
properly balanced. If we have too many $\Wstate$ sites, we will eventually
reach by forward transitions a configuration that looks
like $\ldots \estate \xup \Ustate \Wstate \Wstate \Estate \ldots$ (with potentially more
$\Wstate$ sites). This configuration will evolve as follows:
\begin{center}
\begin{tabular}{c}
\estate \xup \Ustate \Wstate \Wstate \Estate  \\
\estate \estate \xra \Wstate \Wstate \Estate  \\
\estate \estate \wstate \xra \Wstate \Estate \\
\estate \estate \wstate \wstate \xra \Estate \\
\estate \estate \wstate \wstate \xdo \Estate \\
\estate \estate \wstate \xla \Wstate \Estate \\
\estate \estate \xla \Wstate \Wstate \Estate \\
\estate \estate \xup \Wstate \Wstate \Estate \\
\end{tabular}
\end{center}
This creates illegal pair \xup \Wstate in item $12$.
Next we consider what happens 
if the number of~ $\Estate$ sites and $\estate$ sites are
properly balanced and we have too many $\Ustate$ sites. We
start with the case where there is a surplus of two or more $\Ustate$ sites:
\begin{center}
\begin{tabular}{c}
\ldots \estate \xup \Ustate \Ustate \Estate \ldots  \\
\ldots \estate \estate \xra \Ustate \Estate \ldots  \\
\ldots \estate \estate \ustate \xra \Estate \ldots  \\
\ldots \estate \estate \ustate \xdo \Estate \ldots  \\
\end{tabular}
\end{center}
This  creates illegal pair \ustate \xdo from item $12$.
Now if there is only one extra $\Ustate$ site:
\begin{center}
\begin{tabular}{c}
\ldots \estate \xup \Ustate  \Estate \ldots  \\
\ldots \estate \estate \xra  \Estate \ldots  \\
\ldots \estate \estate  \xdo \Estate \ldots  \\
\end{tabular}
\end{center}
Once again, this creates illegal pair \estate \xdo from item $12$.

Finally, we handle the case where we have a double arrow
instead of a single arrow (or vice versa).
In the up-arrow case,  $\leftend \xup$ and $\estate \xdup$ are
both illegal pairs from item 11, so the condition is checked locally.
In the right-arrow case, if there is a configuration with a ~$\xdra$
and more than one $\estate$, it will evolve by reverse transitions
to $\estate \estate \xdra \ldots$ which transitions to $\estate \xdup \Ustate \ldots$
which contains illegal pair \estate\xdup from item 11.
Similarly, a state with one $\estate$ and a control in state $\xra$
will transition in reverse to $\leftend \estate \xra \ldots$ which 
will go to $\leftend \xup \Ustate \ldots$ which contains illegal pair \leftend\xup from item 11.

\end{proof}

\begin{lemma}
\label{lem:consistent}
If a state in the standard basis is bracketed and legal but not consistent, it will evolve eventually 
(via forward or backwards rules) to a configuration which is not legal.
\end{lemma}

\begin{proof}
Since the state is bracketed, we know that if it is not balanced, then it will evolve to
an illegal state, so we can assume that the state is balanced but  not consistent.
This means that there must be a pair of~ $\Estate$ and $\estate$ particles
that don't have the same qubit but should.
Eventually, we will transition backwards to this pair:
\begin{center}
\begin{tabular}{c}
$\ldots \estate_0 \ustate \ldots \ustate \wstate \ldots \wstate \xla \Estate_1 \Estate \ldots$   \\
$\ldots \estate_0 \ustate \ldots \ustate \wstate \ldots \wstate \wstate \xdo_1 \Estate \ldots $  \\
$\ldots \estate_0 \ustate \ldots \ustate \wstate \ldots \wstate \wstate \xra_1 \Estate \ldots $  \\

$\ldots$ \\
$\ldots \estate_0 \xra_1 \Ustate \ldots \Ustate \Wstate \ldots \Wstate \Wstate \Estate \ldots $  \\
\end{tabular}
\end{center}
This creates  a violation with one of the constraints in item $13$. The result would be similar if
the control states was $\xdra$ or the differing bits were swapped.
\end{proof}

Now that we have characterized the paths in the state graph that
are composed of legal configurations, we need to bound the 
spectral gap of 
$H_{trans}+H_{legal}$. We first need to bound the length of the paths.

\begin{lemma}
\label{lem:chain-length}
The length of any chain of well-formed states in the state graph is
at most $n^2$, where $n$ is the number of particles in the chain.
\end{lemma}

\begin{proof}
We associate an ordered pair $(x,y)$ with
every well-formed configuration, where $x$ is
the number of sites in a $\estate$ or a $\Estate$ state.
If the control site is in a state $\xup$, $\xdo$ or $\xdup$, then $y=n$.
If the control state is in state $\xra$ or $\xdra$,  then $y$ is the
number of sites in state $\ustate$ or $\wstate$ that are to the left of
the control state. If the control state is in state $\xla$, then
$y$ is the
number of sites in state $\Ustate$ or $\Wstate$ that are to the right of
the control state. 
We define an ordering on these pairs by first comparing the first index.
If the first index is the same, we compare the second index.
It can be easily verified that if a transition rule applies to a 
configuration in the forward direction, the new configuration is associated with a  pair
of strictly greater value. Similarly, reverse transitions take a configuration
to a configuration associated with a pair
of strictly lesser value. Since there are at most $n^2$ possible pairs,
the lemma follows.
\end{proof}

Let $\calS_p$ denote the space spanned by the basis states within a
path $p$. 
Note that $\calS_p$ is closed under $H_{trans}$ and $H_{legal}$.
$H_{legal}$ when restricted to $\calS_p$ and expressed in the
standard basis is diagonal with non-negative integers along the
diagonal.
$H_{trans}$ when restricted to $\calS_p$ and expressed in the
standard basis has the form:
\[
\left(
\begin{array}{rrrrrrr}
\smfrac{1}{2} & \mns \smfrac{1}{2} &0 & & \cdots& & 0 \\ \mns \smfrac{1}{2} & 1 & \mns \smfrac{1}{2} & 0 &
\ddots & & \vdots\\ 0 & \mns \smfrac{1}{2} & 1 & \mns \smfrac{1}{2} & 0 & \ddots & \vdots\\ & \ddots & \ddots
& \ddots & \ddots & \ddots & \\ \vdots& & 0 & \mns \smfrac{1}{2} &1 & \mns \smfrac{1}{2}& 0 \\ & & & 0 & \mns
\smfrac{1}{2} &1 & \mns \smfrac{1}{2} \\ 0& & \cdots& & 0&
\mns \smfrac{1}{2} & \smfrac{1}{2} \\
\end{array}
\right)
\]

We can now invoke Lemma~$14.4$ from \cite{Kitaev:book} to lower
bound the energy of the overall Hamiltonian for a subspace
$\calS_p$ corresponding to a path with at least one illegal state.

\begin{lemma}
\label{lemma:Kitaev} Let $A_1$, $A_2$ be nonnegative operators, and
$L_1$, $L_2$ their null subspaces, where $L_1 \cap L_2 = \{0\}$.
Suppose further that no nonzero eigenvalue of $A_1$ or $A_2$ is
smaller than $v$. Then
\begin{equation*}
A_1 + A_2 \geq v \cdot 2 \sin^2 {\theta/2},
\end{equation*}
where $\theta = \theta (L_1, L_2)$ is the angle between $L_1$ and
$L_2$.
\end{lemma}

In our case, $A_1$ is the propagation Hamiltonian $H_{trans}$, and its
null state, restricted to $\calS_p$, is the equal superposition
over all states in the path $p$. 
 $A_2$ is the  Hamiltonian $H_{legal}$, diagonal in
the standard basis. Then $\sin^2 \theta$ is the 
fraction of illegal states in the path.  
The minimum nonzero eigenvalue of $H_{legal}$ is
$1$, but (as in~\cite{Kitaev:book}) the minimum nonzero eigenvalue
of $H_{trans}$ is $\Omega(1/K^2)$. In our case $K$, is the length of the 
path which by Lemma \ref{lem:chain-length} is  $O(n^2)$.
Thus, if $p$ is a path containing an
illegal state, all
states in $\calS_p$ have an energy at least $\Omega(1/K^3) = \Omega(1/n^6)$.

Before we summarize the results of this section, we will define a set of
states which we will use to characterize the ground space of
$H_{trans} + H_{legal}$.  For each $x \in \{0,1,\}^m$,
we define $\ket{ \phi_x}$ to be the uniform superposition of the states in
the path that begin with the state in configuration 
~$\leftend \xdup \Ustate^m \Wstate^m \rightend$ whose qubit values
in the $\Ustate$ particles are set according to $x$. 
\begin{lemma}
\label{lem:trans-legal}
Consider  the set of bracketed, well-formed states. Let $\calS$
be the space spanned by these states.
If $n$ is even, then the ground energy of $(H_{trans} + H_{legal}) |_{\calS}$
is $\Omega(1/n^6)$. 
If $n$ is odd, the ground energy is $0$, the spectral gap is $\Omega(1/n^6)$
and the null space is spanned be the $\ket{\phi_x}$.
\end{lemma}

\begin{proof}
Consider a path in the configuration graph consisting of well-formed,
bracketed states. $H_{trans} + H_{legal}$ is closed on the space 
spanned by the states in the path.
If there is a state in the path which is balanced and 
consistent, then by definition $n$ must be odd. Furthermore, we know 
by Lemma \ref{lem:goodstates} that the initial state in the path is a good start state and
that the path contains no illegal states. The uniform superposition
of all states in this path is an eigenstate of $H_{trans} + H_{legal}$ with
zero energy.

If there is a state in the path which is either not balanced or not consistent,
then by Lemmas \ref{lem:balanced} and \ref{lem:consistent},
the path must contain an illegal state. Since the length of any path is
at most $n^2$, the lowest eigenvalue in the subspace spanned by the states
in this path is  $\Omega(1/n^6)$.
\end{proof}

\subsection{Initializing Qubits}
\label{sec:init}

We now add another term to each of the particles which will force the ground state
to be a highly entangled state. This term  is $\ketbra{\Ustate_-}{\Ustate_-}$.
Expressing $\ket{\Ustate_-}$ in terms of standard basis states, we get that $\ket{\Ustate_-} = \frac{1}{\sqrt 2}
(\ket{\Ustate_0} - \ket{\Ustate_1})$. Thus,
$$\ketbra{\Ustate_-}{\Ustate_-} = \frac 1 2 ( \ketbra{\Ustate_0}{\Ustate_0} - \ketbra{\Ustate_0}{\Ustate_1} 
- \ketbra{\Ustate_1}{\Ustate_0} + \ketbra{\Ustate_1}{\Ustate_1} ).$$
$H_{init}$ is the Hamiltonian obtained from summing this term
as applied to all particles in the chain. 
Define 
$$\ket{\phi_g} = \frac{1}{2^{\frac m 2 }}\sum_{x \in \{0,1\}^m} \ket{\phi_x}.$$

\begin{lemma}
\label{lem:trans-legal-init}
Consider a quantum system consisting of
a chain of $n$ particles, where $n$ is odd.
Let $\calS$ be the space spanned by well-formed bracketed 
standard basis states.
$H_{trans}+H_{legal}+H_{init}$ restricted to $\calS$
has a spectral gap of $\Omega(1/n^6)$ and $\ket{\phi_g}$ is
its unique zero energy state.
\end{lemma}

\begin{proof}
Since $H_{init}$ is non-negative, any state in $\calS$ outside
the space spanned by the $\ket{\phi_x}$ will have energy at
least $\Omega(1/n^6)$ by Lemma \ref{lem:trans-legal}.
The space spanned by the $\ket{\phi_x}$ is also spanned by
a different basis: $\ket{\phi_a}$, where $a \in \{+,-\}^m$ and
$\ket{\phi_a}$ is the uniform superposition of all states in the path
whose starting state is the state in configuration
 $\leftend \xdup \Ustate^m \Wstate^m \rightend$
with the qubits in the $\Ustate$ sites set according to $a$.
The $\ket{\phi_a}$ are all zero eigenstates of $H_{trans}+H_{legal}$.
Each $\ket{\phi_a}$ is also an eigenstate of $H_{init}$. 
The only $\ket{\phi_a}$ for which  $H_{init} \ket{\phi_a}=0$
has  $a = \ket{+}^m$ (which is
exactly $\ket{\phi_g}$ ). 

Now consider some  $\ket{\phi_a}$ with $a \neq \ket{+}^m$.
This state will violate $H_{init}$ in at least
one term for at least the first state in the path. Since the path
has length at most $n^2$, we know that $\bra{\phi_a} H_{init} \ket{\phi_a} \ge 1/n^2$.
Thus, the energy penalty of $H_{init}$ for $\ket{\phi_a}$  is at least $1/n^2$.
\end{proof}

\subsection{Boundary Conditions}

We now want to add an energy term
that will penalize states that are not bracketed.
If we can use a position-dependent term on the first and
the last particles in our chain, we could simply add the term
$(I - \ketbra{\leftend}{\leftend}- \ketbra{\rightend}{\rightend})$
to the endpoints.
This would add a penalty of at least one to any well-formed state which is not
bracketed. The resulting Hamiltonian is frustration free.

Alternatively, we can apply  the same
term to every particle.
$H_{bracket}$ is the Hamiltonian obtained from summing this term
as applied to all particles in the chain. 
In order to do this, we need to weight $H_{trans}+H_{legal}+H_{init}$
to ensure that we don't have endmarkers occurring in the middle
of the chain.

\begin{lemma}
Let  $H = 3 (H_{trans}+H_{legal}+H_{init}) + H_{bracket}$,
the unique ground state of $H$
is $\ket{\phi_g}$ and its spectral gap 
is $\Omega(1/n^7)$.
\end{lemma}

\begin{proof}
Let $\calS$ be the space spanned by the set of states in the
standard basis that are well-formed and bracketed. 
$H$ is closed on $\calS$. 
First we consider standard basis states outside of $\calS$.
If the state is not well-formed, it will have energy at least $3$
from the $3 H_{legal}$ term. Let $s$ be the number of particles
in a state $\rightend$ or $\leftend$ that are not 
one of the endpoints of
the chain. Each one of these particles participates in at least one
illegal pair and therefore contributes at least $3/2$ to the 
total energy from the $H_{legal}$ term. Therefore the state has
a total cost of at least $3 \max\{s/2,1\}$ from the $H_{legal}$ term.
The energy from $H_{bracket}$ is
at least $n-2-s$. 
Since $H_{trans}$ and $H_{init}$ are both non-negative,
the energy is at least $n-2-s + 3 \max\{s/2,1\} \ge n-1$ for any standard basis state that is not 
well-formed. If a standard basis state is well-formed but not bracketed,
it will have at most one $\leftend$ or $\rightend$ site. This comes
from our characterization of well-formed states in Lemma \ref{lem:well-formed}.
Thus, it will have energy of at least $n-1$ from the $H_{bracket}$ term.

Any state in $\calS$ is an eigenstate of $H_{bracket}$ with
eigenvalue $n-2$.
The ground state of $H$ is still  $\ket{\phi_g}$ but now with energy
$n-2$ instead of $0$. Any other eigenstate
in $\calS$ has energy that is  $\Omega(1/n^6)$ from the 
$3(H_{trans}+H_{legal}+H_{init})$ term which will give an overall
energy of $(n-2) + \Omega(1/n^6)$.
Note that $\lVert H \rVert$ is $O(n)$. This comes from the observation that
$H$ has energy $O(1)$ for each particle or pair of particles and there are
$n-1$ neighboring pairs in the system.
$H$ can then be normalized so that $\lVert H \rVert$ is $O(1)$ which will give a 
spectral gap of $\Omega(1/n^6)$.
\end{proof}

\subsection{Entropy of Entanglement}
\label{sec:entropy}

We will use the following lemma several times in our discussion of the 
entanglement in the finite chain
in this section and the discussion of the cycle in the next section. 

\begin{lemma}
\label{lem:ent}
Let $\ket{\psi_i}$ for $1 \le i \le r$ be a set of
states of a
quantum system of $n$ particles.  Let $A$ be a subset of the particles
and let $B$ be the complement of $A$.
For each state $\ket{\psi_i}$, let $S_i$ be the set of
standard basis states in the support of
$\ket{\psi_i}$ and let $S_i^A$ be the
resulting set when each state in $S_i$ is traced down to the
particles in $A$. $S_i^B$ is the set resulting from tracing down
the states in $S_i$ to the particles in $B$. $\rho_i$ is the density matrix for $\ket{\psi_i}$
and $\rho_i^A$ is the resulting state when $\rho_i$ is traced down to
the particles in $A$. Define a new state
$\ket{\psi} = \sum_{i=1}^r \alpha_i \ket{\psi_i}$.
If all the $S_i^A$ are mutually disjoint or all the $S_i^B$ are mutually disjoint, then
$$S(\rho^A) \ge \sum_{i=1}^r | \alpha_i |^2 S(\rho^A_i).$$
\end{lemma}

\begin{proof}
Let's assume first that the $S_i^B$ sets are mutually disjoint.
We will establish that 
$\rho^A = \sum_{i=1}^r | \alpha_i |^2 \rho^A_i$.
The lemma then follows from the fact that the entropy is
concave. 
$$\rho = \sum_{i=1}^r \sum_{j=1}^r \alpha_j^* \alpha_i \ketbra{\psi_j}{\psi_i}
= \sum_{i=1}^r |\alpha_i|^2 \rho_i + \sum_{i \neq j} \alpha_j^* \alpha_i \ketbra{\psi_j}{\psi_i}.$$
The last sum consists of terms which are in turn sums over terms of the form 
$c \ketbra{x}{y}$, where $c$ is a complex number,
$x \in S_j$ and $y \in S_i$ for $i \neq j$. We can express
 $x$ as $a_x b_x$ where $a_x \in S_j^A$ and
$b_x \in S_j^B$. Similarly, we can express $y$ as
$a_y b_y$ where $a_y \in S_i^A$ and
$b_y \in S_i^B$. 
When we trace out the particles in $B$, the term $c \ketbra{x}{y}$
becomes $c \ketbra{a_x}{a_y} \braket{b_x}{b_y}$. By assumption,
$b_x \neq b_y$, so all of the terms in $\ketbra{\psi_j}{\psi_i}$
go to zero when $i \neq j$.

If we know that the $S_i^A$ sets are mutually
disjoint, we can apply the result to the set $B$ and use the fact that
$S(\rho^A_j) = S(\rho^B_j)$ for all $j$ and $S(\rho^A) = S(\rho^B)$.
\end{proof}

Now we need to determine the entropy of entanglement for the ground state
$\ket{\phi_g}$. We start by calculating the number of configurations in a path
that begins with a good start state.
We define an {\em iteration} to be the sequence of configurations beginning with
the control particle in state~ $\xup$ or $\xdup$ until it transitions to
$\xup$ again. The first configuration in the path has a ~$\xdup$
control state and the last has a $\xup$ control
state. If there are $m$ particles in state $\Ustate$ at the beginning
of an iteration, the iteration takes $4m+1$ transitions. 
$m$ ranges from $(n-3)/2$ down to $1$
which gives $(n-3)^2/2 + 3(n-3)/2$ transitions
and $T = (n-3)^2/2 + 3(n-3)/2 +1$ configurations in the path.

We will need to divide the path into two parts since only the latter part of the
path has high entanglement. We break the path at the point when the state
has $(n-3)/4+1$ particles in state $\estate$.
Let $T_1$ denote the number of configurations in the first part of the path
and $T_2$ the number of configurations in the second part of the path.
The second part of the path corresponds to the last $(n-3)/4$ iterations and so
$T_2 = (n-3)^2 / 8 + 3(n-3)/4 +1$. For every $n\ge 5$, there is some
constant $c \ge 1/4$ such that $c T_2 = T$.
Let $\ket{\phi_1}$ denote a uniform superposition of the first $T_1$
configurations in the path and $\ket{\phi_2}$ the last $T_2$ configurations
in the path.
Recall that each configuration corresponds to a state which is a superposition
of the $2^m$ basis states corresponding to the $2^m$ ways of setting the
qubits in the 2 dimensional subsystems. Even if there are more than $m$
particles in states that can hold a qubit, we know that there are only
$2^m$ ways to set the values of the qubits since we are guaranteed that
the state is consistent (i.e. entangled pairs are really entangled).
We have that 
$$ \ket{\phi_g} = \sqrt{(1-c)} \ket{\phi_1} + \sqrt{c} \ket{\phi_2} ,$$
where $\braket{\phi_1}{\phi_2} = 0$.
All of the configurations in 
  $\ket{ \phi_2}$ start with $\leftend \estate^{s+1} \ldots$, where $s= (n-3)/4$.
The configurations in the support of $\ket{ \phi_1 }$ have at most
$s$ particles in state $\estate$. 
This means that when we trace out at most
$n-s-2$ 
particles on the right
end of the chain, we can invoke 
Lemma~ \ref{lem:ent}.
Thus, we can lower bound the entropy of entanglement for $\ket{\phi_2}$
which will serve to lower bound the entropy of entanglement for $\ket{\phi_g}$
to within a constant factor.
Note that if $s < (n-3)/4$ and we
trace out $n-s-2$ particles, we need to break the
path at the place where there are $s+1$ particles in state
~$\estate$, but the latter portion of the path will
be larger and this will only serve to increase the value of
$c$.

$\ket{\phi_2}$ is a uniform superposition of states in the standard basis.
We can organize these into $2^{s}$ equally sized sets corresponding to the
value of the qubits in the first $s$ particles in state ~$\estate$.
Since these first $s$ qubits are entangled with qubits on the right
end of the chain, if we take two standard basis states from two 
different sets, these states must also differ somewhere in their
last $s$ sites. Thus if we trace out $t$ particles 
on the right end of the chain
for any $t \in \{s+2, n-s-2\}$, the resulting reduced density matrix
expressed in the standard basis will be block diagonal with 
$2^{s}$ blocks each of which has a trace of $2^{-s}$.
The entropy of the reduced density matrix is therefore at least $s$.

\section{Cycles and the Infinite Chain}

We now describe how to extend the construction for finite chains to cycles
and the infinite chain.
The parameter $n$ is no longer the number of particles in the system but
just a parameter of the Hamiltonian that determines the 
spectral gap and a bound on the entanglement
entropy in the ground state. We will assume throughout
this section that $n$ is odd and that the number of particles in the cycle will be
$nt$ for any $t \ge 2$. We will show bounds on the spectral gap and the entanglement entropy
that are independent of $t$, so as $t$ goes towards infinity, the bounds will
still hold which implies  that they also hold for the infinite chain.
The
ground state is degenerate but any state in the ground space will
exhibit entanglement entropy that is linear in $n$.
As before, we describe a single two-particle term
and apply that term to every neighboring pair on the cycle.

$H_{trans}$ remains unchanged, but
we make several small changes to the Hamiltonian $H_{legal}$. The first
change is that
we allow the pair $\rightend \leftend$. 
For a particular state, we will refer to a sequence of sites extending
from a  $\leftend$ site through the next  $\rightend$ site as
a {\textbf segment}. 
The set of legal and well-formed
states is exactly the same as it was for the finite chain
except that we can now have
more than one segment around the cycle.
For example, we could have
the following state wrapped around a cycle:
$$\underbrace{\leftend \estate \estate  \ustate \xra \Ustate \Wstate \Wstate \Estate \Estate \rightend}_{\mbox{Segment~} 1}
\underbrace{\leftend \xdup \Ustate \Ustate \Ustate \Wstate \Wstate \Wstate \rightend}_{\mbox{Segment~} 2}
\underbrace{\leftend \estate \xup \Estate \rightend}_{\mbox{Segment~} 3}.$$
Note that it would be possible to replace the pair $\rightend \leftend$
by a single delimiting symbol, but it will be convenient to
use the same notation we have developed in the previous section.

We will also add some additional illegal pairs.
These are $\xdup \rightend$ and anything of the form
$\leftend X$ for any state $X$ not equal to $\xdup$ or
$\estate$.
These additional illegal pairs serve to
disallow segments of length two or three
because a sequence of the form $\leftend X \rightend$ or $\leftend \rightend$ 
will contain
an illegal pair. (The pair
$\estate \rightend$ is already disallowed in the original list
of illegal pairs in item $4$.) They have no effect on the ground state
of $H_{trans} + H_{legal}$ for larger chains or segments
since they never appear in the
ground state configurations. 

If a standard basis state is well-formed then every occurrence of ~$\rightend$
has a $\leftend$ to its immediate right and every occurrence of $\leftend$
has a $\rightend$ it its immediate left. Thus, we can assume that a standard basis
state in the support of a ground state can be divided into valid segements. Of course, it
is possible that there are no segments in which case the state could simply be a
 letter state (e.g., $\estate$ or $\Wstate$) repeated
around the entire cycle. Later in this section we will introduce a term that will
be energetically favorable to standard basis states that have at least one segment,
so we will focus for now on those well-formed states that have at least one segment.
Fix a set of locations for the $\rightend \leftend$ pairs in the cycle, which will
then determine the segments.  Let $\calS$ be the
subspace spanned by all well-formed states in the standard basis
that have these segments.
$H_{trans}$ is closed over $\calS$ as it was for the chain.
The final Hamiltonian $H$
will be the sum of $H_{trans}$ and a set of terms which
are all diagonal in the standard basis which means that
$\calS$ will also be closed under $H$ as well.
We will characterize the
eigenstates and corresponding eigenvalues
of $H$ in $\calS$.

Define $H_{chain} = H_{trans} + H_{legal} + H_{init}$.  These
are the terms that we borrow from the previous section on
1D  chains (with the changes to $H_{legal}$
mentioned above).
We will add in another Hamiltonian $H_{size}$ that will 
 be designed to be energetically favorable to segments of size $n$.
The final
Hamiltonian $H$ will have the form $p(n) H_{chain} + H_{size}$
for some polynomial in $n$.

Since all two-particle terms are zero on the pair $\rightend \leftend$,
we can omit the two-particle terms which span two segments when considering 
$H|_{\calS}$. Now $H$ can be divided into a sum of
terms, each of which acts on particles entirely within a segment.
Let $H^i$ be the terms which act on particles within segment $i$.
We can define $H^i_{size}$ and $H^i_{chain}$ similarly.
An eigenstate of $H$ in $\calS$ is then a tensor product of eigenstates
of each $H^i$ acting on the particles in segement $i$. The energy is the sum of the energies
of each $H^i$ on their corresponding eigenstate.

Consider $H_{trans} + H_{legal} + H_{init}$ from the previous section
restricted to the subspace spanned by the
set of all well-formed bracketed states
acting on a chain of length $l$.
This is exactly the same operator as
$H^i_{chain}$ restricted to 
subspace $\calS$, acting
on the particles in a segment $i$ of length $l$
(with all other particles traced out).
From Lemma \ref{lem:trans-legal-init}, we know that 
if $l$ is odd, the spectral gap is $\Omega(1/l^6)$
and there is a unique zero eigenstate $\ket{\phi_g^l}$
(with an additional parameter $l$ now
denoting the length of the chain).
If $l$ is even, then the minimum energy is $\Omega(1/l^6)$.

We are now ready to define the final component of $H$.
Recall that $T_n$ is the length of the path in the state graph corresponding to
$\ket{\phi_g^n}$. In other words, $T_n$ is the number of configurations
in the support of $\ket{\phi_g^n}$.
In Section \ref{sec:entropy}, we determined that $T_n = (n-3)^2/2 + 3(n-3)/2 + 1$.
$$H_{size} = \frac{1}{n} I - 2
\ketbra{\leftend}{\leftend} + \frac{T_n}{n-2} \left( \ketbra{\xdo}{\xdo} +
\ketbra{\xup}{\xup} + \ketbra{\xdup}{\xdup}\right).$$
We will analyze the ground energy of a segment as a function of its length.
We will need
to use the Projection Lemma from \cite{KempeKitaevRegev06} 
which will allow us to focus on the
ground space of $H^i_{chain}$. 

\begin{lemma}
\label{lem:proj}
\cite{KempeKitaevRegev06} 
Let $H = H_1 + H_2$ be the sum of two Hamiltonians acting on a Hilbert space
$\calH = \calT + \calT^{\perp}$. The Hamiltonian $H_2$ is such that
$\calT$ is a zero eigenspace for $H_2$ and the eigenvectors in $\calT^{\perp}$
have value at least $J > 2 \lVert H_1 \rVert$. Then
$$\lambda(H_1 |_{\calT}) - \frac{ \lVert H_1 \rVert^2}{J - 2 \lVert H_1 \rVert}
\le \lambda(H) \le \lambda(H_1 |_{\calT}).$$
\end{lemma}

\begin{coro}
\label{cor:proj}
There is a polynomial $p(n)$ such that $p(n)$ is $O(n^{10})$ and for any
segment of size $l \le 2n$ and
$H^i = p(n) H^i_{chain} + H^i_{size}$, 
$$\lambda(H^i|_{\calS}) \ge \bra{\phi_g^l} H^i_{size} \ket{\phi_g^l}  - 1/2 n^2.$$
\end{coro}

\begin{proof}
We use the projection lemma with
$H_2 = p(n) H^i_{chain}$ and $H_1 = H^i_{size}$. Note that $H_1$ need not
be positive, although it does need to be positive on
$\calT$ in order to yield a non-trivial lower bound. 
$\calT$, the ground space for $H^i_{chain}$, is just the state
$\ket{\phi_g^l}$.
We need to establish that $\lVert H^i_{size} \rVert = O(n)$.
Since $l \le 2n$, the first term is $O(n)$.
The Hilbert space $\calS$ is the set of all well-formed, bracketed states for that
segement, so there can be at most one site in $\xdo$, $\xup$ or $\xdup$ and
at most one site in $\leftend$. Thus the second two terms in $H^i_{size}$
are at most $1 + T_n/n$ for any state in $\calS$
which is also $O(n)$.
The spectral gap of $H^i_{chain}$ is $\Omega(1/n^6)$, so we can choose
 $p(n)$ so that $p(n)$ is $O(n^{10})$
and $J$ (the spectral gap of $p(n) H^i_{chain}$) is at least
$2 n^2 \lVert H_1 \rVert^2 + 2 \lVert H_1 \rVert$. Using Lemma \ref{lem:proj}, we
can lower bound $\lambda(H^i|_{\calS})$ by $\bra{\phi_g} H^i_{size} \ket{\phi_g}   - 1/2 n^2$.
\end{proof}

Note that we are not able to use the projection lemma for very large $l$ because the
$\Omega(1/l^6)$ gap will not be large enough. In the lemma below, we 
determine the ground energy of a segment 
as a function of its length. Large $l$ (greater than $2n$) are dealt with
separately with an argument that does not require the projection lemma.

\begin{lemma}
\label{lem:size}
The operator $H^i$ acting on the $l$
particles of segment $i$
restricted to well-formed bracketed states will
have ground energy $0$ and spectral gap $\Omega(1)$ for $l=n$.
The ground
energy is at least $1/2n^2$ for any other value of $l$.
\end{lemma}

\begin{proof}
Any sequence $\leftend X \rightend$ will have an illegal pair.
$\leftend \rightend$ is also illegal. Therefore, we can assume that
$l \ge 4$.
We consider four different cases based on the size of the segment $l$.
\begin{description}
\item { $\mathbf {l=n}$:}

Consider the state $\ket{\phi_g^n}$. $\bra{\phi_g^n} H_{chain} \ket{\phi_g^n} = 0$.
Recall that $\ket{\phi_g^n}$ is a uniform superposition of states.
There are $T_n$ distinct configurations represented in the support of
$\ket{\phi_g}$ each of which as $2^m$ states for $m = (n-3)/2$.
Each configuration has  one has a $\rightend$ site and the number of configurations
that contain
a $\xdo$, $\xup$ or $\xdup$ state is $n-2$.
Therefore
$$\bra{\phi_g^n} H^i_{size} \ket{\phi_g^n} = \frac n n - 2 + 
\frac{T_n}{(n-2)} \frac{(n-2)}{T_n} = 0.$$
Any state $\ket{\psi}$ that is orthogonal to $\ket{\phi_g^n}$
and is also in the subspace spanned by the well-formed
states for segments of length $n$ 
will have 
$\bra{\psi} p(n) H_{chain} \ket{\psi} \ge n^4$ and 
$\bra{\psi} H_{size} \ket{\psi} \ge -2n$. Thus, the spectral gap of $H^i$ will be $\Omega(1)$.
\item { $\mathbf {l > 2n}$:}

Let $\psi$ be a state in the standard basis that is well-formed, bracketed
and has length $l$.
We will only lower bound $\bra{\psi} H^i_{size} \ket{\psi}$. Since
$H^i_{chain}$ is non-negative, the lower bound will hold for all
of $H^i$. Furthermore, we will omit the last term in $H_{size}$ because this
only adds to the energy.
Every standard basis state in a bracketed well-formed segment of length $l$ has
exactly one occurrence of ~~$\leftend$. Therefore the energy of a segment of length
$l$ will be at least $l/n - 2$. Since $l > 2n+1$, this will be at least
$1/n$.
\item {$\mathbf {2n \ge l > n}$:}

We will first handle the case that $l$ is even. From Lemma \ref{lem:trans-legal},
we know that the lowest eigenvalue of $H_{trans}+H_{legal}$ on
a chain of length $l$ is $\Omega(1/l^6)$ which is in turn $\Omega(1/n^6)$.
The other terms in $H^i$ are positive and the $H_{trans}+H_{legal}$
are weighted by a factor of $p(n)$ which bring the lowest energy to $\Omega(1)$.

Since we can assume that $l$ and $n$ are both odd, we know that $l \ge n+2$.
We
 will use the projection lemma for this case and show that
$\bra{\phi_g^l}H^i_{size}\ket{\phi_g^l} \ge 1/n^2$ which by
Corollary \ref{cor:proj}
will be enough to lower bound $\lambda(H^i)$ by
$1/2 n^2$. 
We osberve that  
$$T_n = \frac{(n-3)^2}{2} + \frac{3(n-3)}{2} + 1 = \frac{(n-1)(n-2)}{2},$$
so $T_n/(n-2) = (n-1)/2$, and
\begin{eqnarray*}
\bra{\phi_g^l}H^i_{size}\ket{\phi_g^l} & = & \frac{l}{n} - 2 
+ \left( \frac{T_n}{n-2} \right) \left( \frac{l-2}{T_l} \right) \\
& = & \frac{l}{n} - 2 + \frac{n-1}{l-1}\\
& = & \frac{l-n}{n} + \frac{n-l}{l-1}\\
& \ge & (l-n)\left( \frac 1 {n} -\frac{1}{l-1} \right) \\
& \ge & (l-n)\left( \frac 1 {n} -\frac{1}{n+1} \right) \ge \frac{2}{n(n+1)} \ge 
\frac{1}{n^2}\\
\end{eqnarray*}
\item {$\mathbf {l < n}$:}

We can use the same reasoning as in the previous case to assume that $l$ is odd.
Since both $l$ and $n$ are  odd, we know that $l \le n-2$.
Now we will use the projection lemma for this case and show that
$\bra{\phi_g^l}H^i_{size}\ket{\phi_g^l} \ge 1/n^2$ which will be enough to lower bound $\lambda(H)$ by
$1/2 n^2$. 
\begin{eqnarray*}
\bra{\phi_g^l}H^i_{size}\ket{\phi_g^l} & = & \frac{l-n}{n} + \frac{n-l}{l-1}\\
& \ge & (n-l)\left( \frac 1 {l-1} -\frac{1}{n} \right) \\
& \ge & (n-l)\left( \frac 1 {n-3} -\frac{1}{n} \right) \ge \frac{2}{n(n-3)} \ge 
\frac{1}{n^2}\\
\end{eqnarray*}
\end{description}
\end{proof}
As a result of Lemma \ref{lem:size}, we know that if the number of particles in the chain
is $nt$, a multiple of $n$, and if $n$ is odd, then the ground energy of $H$ is
zero and the ground space is spanned by the states that consist of $t$ copies of
$\ket{\phi_g^n}$ tensored together. There are actually $n$ such states, which can be obtained
by taking one and rotating it by one position $n$ times along the cycle.
We will call these $\ket{\psi_0}, \ldots , \ket{\psi_{n-1}}$.
Any eigenstate of $H$ that has a segment which is not equal to $n$ will have energy
at least $\Omega(1/n^2)$ while the norm of a single term in $H$ which acts on a pair
of neighboring states is at most $O(n^{10})$. This means the final spectral gap 
is $O(n^{-12})$
when the Hamiltonian for the entire chain is scaled to $O(n)$.
We still need to handle the case where there is a configuration which is
well-formed but has no segments. This would just correspond
to a configuration of all lower case states or all upper case states.
$H_{chain}$ would evaluate to zero on such a state but $H_{size}$ would be at
least $t$.
Since these bounds are independent of $t$ and hold for arbitrarily large $t$,
they hold as $t$ tends towards infinity.

\subsection{Entropy of Entanglement}

Consider the cycle with $nt$ particles,
a basis state $\ket{\psi_i}$ and a set $A$ of contiguous particles
in the cycle. 
We say that a particle in $A$ is {\em good} for $\ket{\psi_i}$ 
if it is the $p^{th}$ particle in a some segment $j$ where
$2 \le p \le n/4$ or $3n/4 \le p \le n-1$ and the
$(n-p)^{th}$ particle in that segment is not in $A$.

\begin{lemma}
\label{lem:good-particles}
Consider a state $\ket{\psi_i}$ and a contiguous set $A$ of $r$ particles
on the cycle.
We assume that $r \le nt-n$.
When $\ket{\psi_i}$ is traced down to the particles in $A$,
the entropy of
the resulting state is at least the number of particles in $A$ that are good for
$\ket{\psi_i}$ divided by $4$.
\end{lemma}

\begin{proof}

The segments in $\ket{\psi_i}$ are fixed and we shall number
them from $1$ to $t$.
$\ket{\psi_i}$ is a tensor product of states $\ket{\psi_i^j}$,
where $\ket{\psi_i^j}$ is the ground state for a finite chain
of length $n$ for segment $j$.
The set of good particles can only come from two different segments. This is because
if a segment contains a good particle, one of the endpoints in $A$ must be contained in that
segment.
We will arbitrarily call these segments $j$ and $k$.
Let $A_j$ be the set of good particles in $j$ and $A_k$ be the set of
good particles in $k$.
The state $\ket{\psi_i}$ can be written as
$\ket{\psi_i} = \ket{\psi_i^j} \otimes \ket{\psi_i^k} \otimes
\ket{\psi_i^{R}}$. Where $\ket{\psi_i^{R}}$ is the state for
the rest of the cycle (all sites not in segment $j$ or $k$).

The support of $\ket{\psi_i^j}$ can be partioned into two sets
depending on whether the good particles are all in an
entangled state ($\estate$ or $\Estate$) or whether there is
a good particle that is not in an entangled state.
Let $\ket{\phi_i^j}$ be the uniform superposition of the states
in which all the good particles are entangled and
$\ket{\hat{\phi}_i^j}$ be uniform superposition of the states
for which there is a good site that is not in an entangled state.

Since the good particles are all either in the first
$n/4$ or last $n/4$  particles in the chain, we can use the same argument 
used in  Section
\ref{sec:entropy} to determine that there is a constant $c_j \ge 1/4$ 
such that a fraction of $c_j$ of the
states in the support of $\ket{\psi_i}$ are in the support of $\ket{\phi_i^j}$.
We can write $\ket{\psi_i^j} = \sqrt{c_j} \ket{\phi_i^j} +
\sqrt{1 - c_j} \ket{\hat{\phi}_i^j}$ and
 $\ket{\psi_i^k} = \sqrt{c_k} \ket{\phi_i^k} +
\sqrt{1 - c_k} \ket{\hat{\phi}_i^k}$. $c_k$ is also at least
$1/4$ although not necessarily equal to $c_j$.

\begin{small}
$$
\ket{\psi_i}  =   \left( \sqrt{ (1 - c_j)(1 - c_k)} \ket{\hat{\phi}_i^j} \ket{\hat{\phi}_i^k} +
\sqrt{ c_j (1 - c_k)} \ket{\phi_i^j} \ket{\hat{\phi}_i^k}  +
\sqrt{  (1 - c_j)c_k} \ket{\hat{\phi}_i^j}\ket{\phi_i^k} +
\sqrt{  c_j c_k} \ket{\phi_i^j} \ket{\phi_i^k} \right) \otimes
 \ket{\psi_i^{R}} .
$$
\end{small}

Furthermore, the four states in the above sum satisfy the conditions for 
for Lemma \ref{lem:ent} for the set $A$.
$\ket{\hat{\phi}_i^j}  \ket{\phi_i^k} \ket{\psi_i^{R}} $
has $|A_k|$ entangled pairs between $A$ and the rest of the cycle.
Similarly, $\ket{\phi_i^j}  \ket{\hat{\phi}_i^k}   \ket{\psi_i^{R}} $
has
$|A_j|$ entangled pairs 
and $\ket{\phi_i^j}  \ket{\phi_i^k} \ket{\psi_i^{R}} $
has $|A_j|+|A_k|$
entangled pairs between $A$ and the
rest of the cycle.
We then have
$$
S(\rho_i^A)  \ge  c_j (1 - c_k) (|A_i|) + (1-c_j) c_k (|A_k|) + 
c_j c_k (|A_j|+ |A_k|) \ge \frac 1 4 (|A_j|+ |A_k|) .$$
\end{proof}

In the next lemma, we extend the lower bound on the entanglement to
an arbitrary superposition of the $\ket{\psi_i}$.

\begin{lemma}
Consider a cycle with $nt$ particles.
Let $\ket{\psi} = \sum_{i=0}^{n-1} \alpha_i \ket{\psi_i}$.
For any fixed $r \le (n-1)t$, 
pick a random set $A$ of $r$ contiguous particles in the cycle.
The expected entropy of entanglement of $\ket{\psi}$
when the state $\ket{\psi}$ is traced down
to $A$ is at least $(\min\{r,n/4\}-2)/16$.
\end{lemma}

\begin{proof}
Consider a particular $\ket{\psi_i}$.
With probability $1/4$, the left end of the segment will fall in the
last $n/4$ particles in a segment. If $r \le n/4$, this means that all
but two of the particles are good (the exceptions
are the sites in state $\rightend$
and $\leftend$). 
If $n/4 < r \le n/2$, then $n/4 -2$ of the particles are good.
With probability $1/4$, the left end will fall in the
range $n/2 + 1$ to $3n/4$. If $r > n/2$, then the number of good
particles is at least $n/4-1$  because $A$ will contain 
all of the particles in the last quarter of the segment. Since $r < nl-n$,
it can not wrap around and contain any of the particles in the first quarter
of that segment.
Thus, with probability at least $1/4$, there are at least $\min\{r,n/4\}-2$
good particles in $A$ for $\ket{\psi_i}$.
Using Lemma \ref{lem:good-particles}, $E[S(\rho^A_i)] \ge (\min\{r,n/4\}-2)/16$.

Let $B$ be the complement of $A$ and
$S_i^B$ be the set of standard basis states in $\ket{\psi_i}$ traced
down to the set $B$.
If $A$ has at most $nt-n$ particles then  every state in every
$S_i^B$ contains a $\leftend$ site.
Furthermore, for the states within a single $S_i^B$, the $\leftend$ sites are the
same and they are all different from the $\leftend$ sites for the states in
a $S_j^B$ for $i \neq j$. Thus, the $S_i^B$'s are all mutually disjoint and
we can  apply Lemma \ref{lem:ent} and linearity of expectations to get
$$E[S(\rho^A) ]\ge \sum_{i=1}^r | \alpha_i |^2 E[S(\rho^A_i)]\ge \frac{\min\{r,n/4\}-2}{16} .$$
\end{proof}

Since the random variable denoting the entropy of entanglement
for a randomly chosen $A$ of size $r$ is in the range
$0$ to $\log(21 \cdot r)$, we can apply Markov's inequality to determine
that with constant probability the entanglement entropy of 
a randomly chosen $A$
is $\Omega(\min\{r,n\})$.

Finally consider the translationally invariant state
$\ket{\Phi} = \sum_{i=0}^{n-1} (1/\sqrt{n}) \ket{\psi_i}$. 
For any fixed set $A$, $A$ will have at least $\min\{r,n/4\}$
good particles for at least $n/4$ of the $\ket{\psi_i}$. 
Applying Lemmas \ref{lem:ent} and \ref{lem:good-particles} to these states,
we get that the
entanglement entropy of $A$ for $\ket{\Phi}$ is
at least $(\min\{r,n/4\}-2)/16 = \Omega(\min\{r,n\})$.

\section{Open Questions}

There still remains an exponential difference in the dependence on $1/\Delta$
between Hastings' area law and the lower bound presented here and that in
\cite{GH08}. Resolving this
discrepancy may involve strengthening the upper bound given in the area law.
There are also issues related to the translationally invariant construction
given here that would be worthwhile to clarify. For example, is it possible
to obtain a construction on the infinite chain that achieves the same
entanglement entropy but with a unique ground state?
Can  one obtain a lower bound of $\Omega(\min\{m,n\})$
for the entanglement entropy on all regions of size $m$ instead of
a constant fraction of the regions?
Is there a 1D Hamiltonian for which the entanglement
is linear
in the region size simultaneously for all region sizes?
The latter property
could only be achieved on a gapless system because the 1D area law
indicates that  any non-zero spectral gap
will imply a finite upper bound on the entanglement entropy for any region.
It is not known whether this can be achieved even for a 
Hamiltonian with position-dependent terms. 
Finally, how robust are the entanglement properties in the
ground state to small fluctuations in the
terms of the Hamiltonian? It seems likely that the construction
presented here will break with small errors in the transition
rules. Is it possible to obtain a fault-tolerant version of
this construction?

\section{Acknowledgements}

The author would like to thank Sergio Boixo, Daniel Gottesman, 
Matthew Hastings, Stephen Jordan and John Preskill for useful discussions.

\bibliographystyle{alpha}
\bibliography{onedimbib}

\end{document}